\documentclass[journal,twoside,web]{ieeecolor}
\usepackage{generic}
\usepackage{cite}
\usepackage{amsmath,amssymb,amsfonts}
\usepackage{algorithmic}
\usepackage{graphicx}
\usepackage{textcomp}
\usepackage[normalem]{ulem}
\usepackage{pifont}
\usepackage{multirow}
\newcommand{\xmark}{\ding{55}}  
\definecolor{lightgray}{rgb}{0.9, 0.9, 0.9}
\usepackage{color}
\usepackage{colortbl}
\usepackage{hyperref}
\hypersetup{hidelinks=true}
\usepackage{textcomp}
\def\BibTeX{{\rm B\kern-.05em{\sc i\kern-.025em b}\kern-.08em
    T\kern-.1667em\lower.7ex\hbox{E}\kern-.125emX}}
\begin{document}
\title{ Diffusion-empowered AutoPrompt  MedSAM }
\author{Peng Huang, 
Shu Hu,    
Bo Peng, 
Xun Gong,
Penghang Yin,
Hongtu Zhu, \IEEEmembership{Fellow, IEEE}, 
Xi Wu,    
and Xin Wang, \IEEEmembership{Senior Member, IEEE} 
\thanks{Peng Huang, Bo Peng, and Xun Gong are with the School of Computing and Artificial Intelligence, Southwest Jiaotong University, Chengdu, China (e-mail: huangpeng@my.swjtu.edu.cn; bpeng@swjtu.edu.cn; gongxun@foxmail.com).}
\thanks{Shu Hu is with the Department of Computer and Information Technology, Purdue University, USA (e-mail: hu968@purdue.edu).}
\thanks{Penghang Yin is with the Department of Mathematics and Statistics, University at Albany, State University of New York, USA (e-mail: pyin@albany.edu).}
\thanks{Hongtu Zhu is with the University of North Carolina at Chapel Hill, USA (e-mail: htzhu@email.unc.edu).}
\thanks{Xi Wu is with the School of Computer Science, Chengdu University of Information Technology, Chengdu, China (e-mail: wuxi@cuit.edu.cn).}
\thanks{Xin Wang is with the College of Integrated Health Sciences and the AI Plus Institute,  University at Albany, State University of New York, USA (e-mail: xwang56@albany.edu).}
\thanks{Corresponding Author: Bo Peng, Xin Wang.}
}

\maketitle

\begin{abstract}
MedSAM, a medical foundation model derived from SAM, has achieved remarkable success across various medical domains. Nonetheless, its clinical adoption encounters two primary challenges: the reliance on labor-intensive manual prompt generation, which places a considerable burden on clinicians, and the lack of semantic labeling in the generated segmentation masks for organs or lesions, which hinders its usability for non-expert users. 
To overcome these limitations, we propose an end-to-end interactive segmentation model, AutoMedSAM. AutoMedSAM employs a diffusion-based class prompt encoder to generate prompt embeddings guided by the prompt class indices and eliminates the reliance on manual prompts. During the diffusion process, the encoder progressively captures the semantic structure and fine-grained features of the target object, injecting semantic information into the prediction pipeline. 
Furthermore, we introduce an uncertainty-aware joint optimization strategy that integrates pixel-based, region-based, and distribution-based losses. This approach harnesses MedSAM’s pre-trained knowledge and various loss functions to enhance the model’s generalization. Experimental evaluations on diverse datasets demonstrate that AutoMedSAM not only achieves superior segmentation performance but also extends its applicability to clinical environments and non-specialist users. 
Code is available at \url{https://github.com/HP-ML/AutoPromptMedSAM.git}.

\end{abstract}

\begin{IEEEkeywords}
MedSAM, medical image foundation model, end-to-end, diffusion model, uncertainty learning
\end{IEEEkeywords}

\section{Introduction}
\label{sec:introduction}
\IEEEPARstart{D}{eep} learning models have traditionally been applied in medicine by designing and training specialized models for specific tasks, achieving significant success~\cite{Azad2024UnetReview, Marinov2024seg, huang2024robustly,wang2024artificial}. 
However, these approaches often require training a model from scratch using corresponding data~\cite{Hatamizadeh2022UNETR, U-Mamba, nnFormer}, leading to low training efficiency and limited transferability to other tasks~\cite{weiss2016survey, hosna2022transfer}. 
Meanwhile, diagnostic-computer interaction is essential in medical scenarios. Ideally, a physician should be able to provide reference information (e.g., an organ or lesion class ID) to the system, which then recognizes and segments relevant targets. However, most existing specialist models do not support the operation of such interactivity.
Recently, prompt-based foundational models in computer vision, such as the Segment Anything Model (SAM)~\cite{SAM}, have demonstrated impressive performance and generalization capabilities in various semantic segmentation tasks based on user-provided prompts~\cite{zhang2023comprehensiveSAM, zhou2024medsam}. 

\begin{figure}[t]
  \centering
  \includegraphics[width=0.95\linewidth]{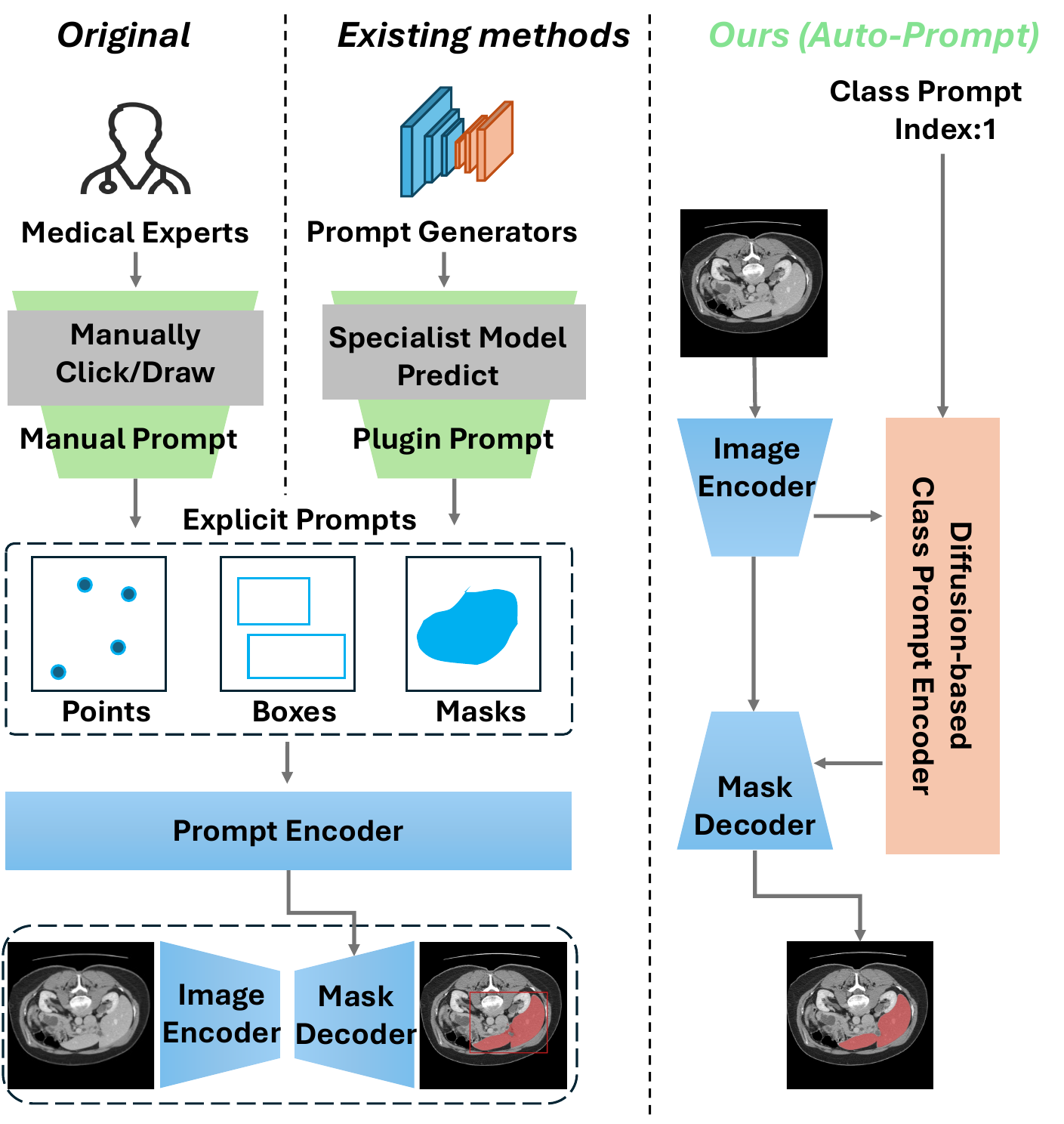}
  \caption{\small Comparison with SAM-based models.(\textbf{Left}) The original SAM model relies on manual prompts from medical experts, restricting its usability and scenarios. (\textbf{Middle}) Current SAM-based methods employ specialist models for prompt generation, but these models are organ- or lesion-specific, limiting SAM's generalizability. (\textbf{Right}) Our method introduces an automatic diffusion-based class prompt encoder, removing the need for explicit prompts, adding semantic labels to masks, and enabling accurate, end-to-end segmentation for non-experts in diverse medical contexts.}
  \label{fig:method_compare}
\end{figure}

Compared to natural images, medical images generally have lower contrast, lower resolution, and high inter-class similarity, with strong domain-specific characteristics. As a result, SAM performs poorly in the medical domain. To address this, Ma et al. proposed the foundational medical model MedSAM~\cite{ma2024medsam}, which has been optimized for the unique characteristics of medical images. In its optimization process, MedSAM utilized over one million medical masks, including common modalities such as MRI and CT, marking the entry of segmentation models in the medical field into the era of large segmentation models.

In busy clinical environments, it is difficult for physicians to provide accurate prompts through labor-intensive manual prompting, but it is a fundamental requirement for obtaining precise masks. On the other hand, for users without medical knowledge, the pixels in the image corresponding to organs or lesions are not immediately intuitive, and the generated segmentation masks lack additional semantic annotations, making them difficult to interpret. This renders the segmentation results of little practical value for non-expert users, thereby limiting the applicability of MedSAM in certain contexts and its potential user base.
Although MedSAM boasts impressive zero-shot and few-shot segmentation capabilities, there are still some challenges.
Specifically: \textbf{(1) Dependency on Manual Prompts}: It requires users to provide precise prompts to segment the target region. However, creating prompts for medical images requires expertise, and in clinical settings, providing explicit prompts like points, bounding boxes, or scribbles is impractical.
\textbf{(2) Limitations of Prompt Precision}: 
The manual nature of these prompts means the error may not be within a controllable range~\cite{li2023autoprosam}. Using the bounding box prompt as an example, MedSAM's performance relies on the discrepancy between the prompt box and the true boundary. However, other targets within the prompt box cannot be avoided, and different organ or lesion categories often exhibit high similarity and low inter-class variation~\cite{yue2024surgicalsam}. 
\textbf{(3) Lack of Semantic Information}: MedSAM cannot obtain the semantic information of the masks it predicts. It can only predict binary masks for each prompt, without associating them with semantic labels~\cite{SAM, ma2024medsam, xie2024masksam}. 


To address these challenges, we propose AutoMedSAM. A comparison between AutoMedSAM and existing methods is shown in Fig.~\ref{fig:method_compare}. We replaced the original prompt encoder with a diffusion-based class prompt encoder. This new encoder uses a lesion or organ index as prompting, incorporating semantic information about the target into the AutoMedSAM learning process. The class prompts and image embeddings are input information to generate prompt embeddings for the mask decoder directly. By generating prompt embeddings directly from class prompts, we eliminate the robustness issues caused by manually provided explicit prompts and transform the semi-automatic MedSAM into a fully automated end-to-end process. Meanwhile, we designed an uncertainty-aware joint optimization training strategy. This strategy can optimize the model training process by combining the advantages of multiple loss functions while transferring the pre-training knowledge from MedSAM to AutoMedSAM. This enables AutoMedSAM to adapt to data of various modalities and effectively extract features of different organs or lesions, enhancing the segmentation performance and robustness of the model.

  

In summary, our contributions are as follows:
\begin{enumerate} 
\item We introduced an interactive end-to-end model, AutoMedSAM, enabling doctors and researchers to rapidly and accurately analyze medical images for more timely diagnosis and treatment, without relying on labor-intensive manual sketching of organs or lesions to guide the segmentation process.
\item We proposed a diffusion-based class prompt encoder that eliminates the need for explicit prompts and facilitates direct learning of latent prompt embeddings from class prompts. Through the diffusion process, the encoder progressively enhances its understanding of the medical targets, thereby improving the effectiveness and generalization capability of the class prompts.
\item We designed an uncertainty-aware joint optimization strategy that combines pixel-based, region-based, and distribution-based loss functions. This approach leverages the pre-trained knowledge of MedSAM and integrates diverse loss components to improve the model’s generalization ability and robustness across multi-modal medical data.

\end{enumerate}

\section{Related Works}

\subsection{SAM-based Medical Image Segmentation}
SAM represents a significant breakthrough in transforming image segmentation from specialized applications to a general-purpose tool~\cite{MAZUROWSKI2023102918}. 
After training on large-scale datasets, SAM builds a broad knowledge base and relies on manually provided explicit prompts with precise locations (e.g., points and bounding boxes) to trigger segmentation responses~\cite{ZHANG2024108238}. However, due to the substantial domain gap between natural and medical images, SAM exhibits limited generalizability in medical imaging. To bridge this gap, studies such as MedSAM and SAM Med2D integrate extensive medical imaging data with specialized fine-tuning strategies, effectively improving performance in medical scenarios~\cite{ma2024medsam, le2024medficientsam, cheng2023sam2d}. However, the accuracy of SAM’s segmentation is sensitive to the positional bias of the prompts (as validated in Section~\ref{sec:manual}). Consequently, the segmentation process often requires extensive manual intervention or professional detectors, turning it into a multi-stage workflow. In certain clinical settings(e.g., surgeries~\cite{yue2024surgicalsam}), providing explicit prompts for every frame is impractical. To address this limitation, methods like MaskSAM and UV-SAM employ task-specific expert models to generate coarse positional prompts, reducing reliance on explicit user input~\cite{xie2024masksam, zhang2024uv}. While these approaches mitigate the need for explicit prompts, they lack fine-grained class-specific prompting capabilities and face scalability challenges~\cite{paranjape2024adaptivesam, yue2023part}. In contrast, our method introduces a class-based prompting approach, leveraging a diffusion process to generate prompt embeddings from image features. This simplified prompting mechanism streamlines the segmentation pipeline, and enhances fine-grained class differentiation.

\begin{figure*}[t]
    \centering
    \includegraphics[width=0.95\textwidth]{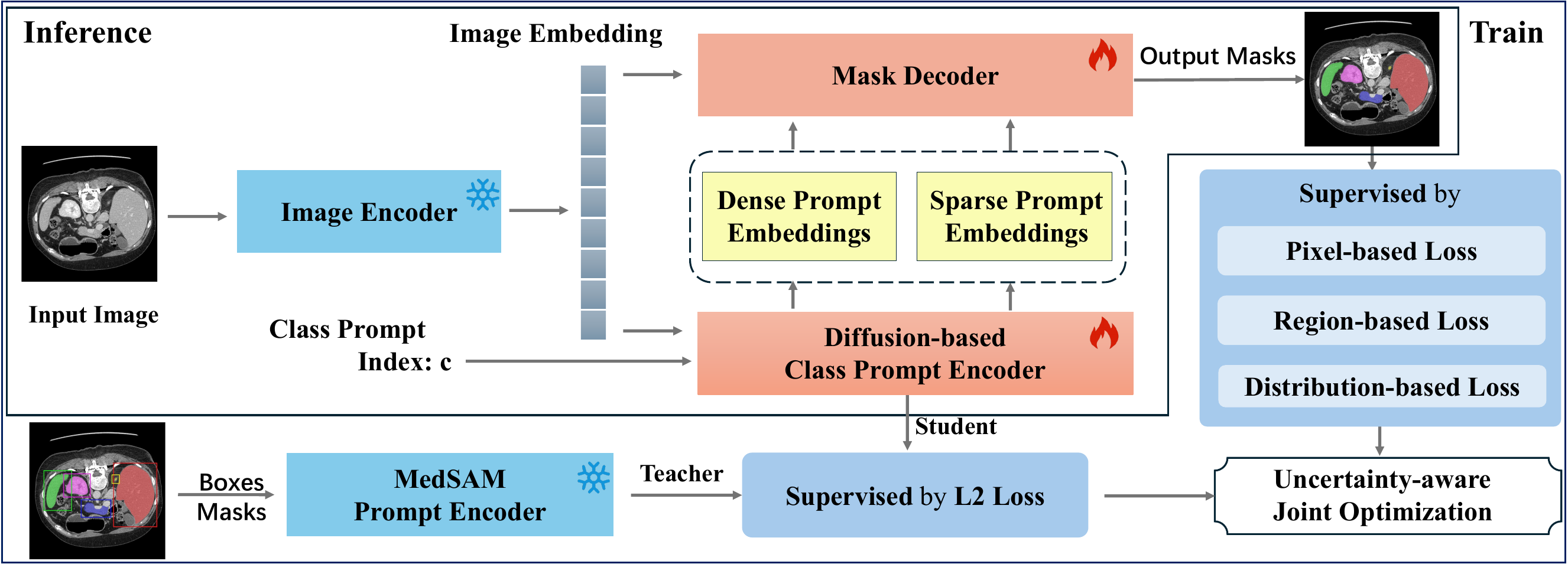}
    \caption{\small An overview of the AutoMedSAM. AutoMedSAM generates dense and sparse prompt embeddings through a diffusion-based class prompt encoder, eliminating the need for explicit prompts. During training, we employ an uncertainty-aware joint optimization strategy with multiple loss functions for supervision, while transferring MedSAM's pre-trained knowledge to AutoMedSAM. This approach improves training efficiency and generalization. With end-to-end inference, AutoMedSAM overcomes SAM's limitations, enhancing usability and expanding its application scope and user base.}
    \label{fig:overview}
\end{figure*}

\subsection{Diffusion Models for Medical Domain}
Diffusion models show strong potential in medical imaging, achieving notable success in tasks such as image generation, segmentation, and classification~\cite{YAN2024112350, zhu2024diffusion, fontanella2024diffusion}. The initial applications of diffusion models in the medical field primarily focused on generating medical data, which has proven useful for medical data augmentation~\cite{kazerouni2023diffusion}. In addition, several scholars have investigated the potential of medical images generated by diffusion models as a substitute for real data in training deep networks. M. Usman Akbar et al. and D. Stojanovski et al. demonstrated that these synthetic data are effective for downstream tasks~\cite{usman2024brain, stojanovski2024efficient}. Recent studies have used diffusion models for cross-modality synthesis~\cite{luo2024target, pan2024synthetic}. For example, DCE-diff addresses data heterogeneity by leveraging multimodal non-contrast images to extract anatomical details from structural MRI sequences and perfusion information from ADC images, enabling the synthesis of early-phase and late-phase DCE-MRI~\cite{ramanarayanan2024dce}. D. Stojanovski et al. demonstrated that visual realism during model training does not necessarily correlate with model performance. Consequently, these models can generate organ or lesion features optimized for deep learning, thereby enhancing the accuracy of downstream tasks. Furthermore, utilizing more efficient models can significantly reduce computational costs~\cite{stojanovski2024efficient}. By progressively refining representations through noise-based generation and denoising, diffusion models inherently capture fine-grained structural details and semantic consistency, which are particularly beneficial for complex medical image tasks.

\section{Method} 

\subsection{Overview of AutoMedSAM}
The optimization of AutoMedSAM is essentially based on the structure of SAM, with the innovative introduction of a diffusion-based class prompt encoder to address the challenges of manual prompts. Specifically, as illustrated in Fig.~\ref{fig:overview}, AutoMedSAM consists of three core modules: an image encoder $E_I$, a diffusion-based class prompt encoder $E_P$, and a mask decoder $D_M$. The input image is denoted as $I \in \mathbb{R}^{h \times w \times 3}$, with spatial resolution $h \times w$. By providing the prompt class $c$, AutoMedSAM can predict the mask corresponding to the class $c$. The image is first processed by the image encoder to generate the image embedding $F_I$. Subsequently, the class prompt encoder $E_P$, based on a diffusion model, processes the image embedding to generate sparse prompt embedding $P_s^{(c)}$ and dense prompt embedding $P_d^{(c)}$ from the target prompts. Finally, the mask decoder combines the image embedding, positional encoding $P_p$, sparse prompt embedding, and dense prompt embedding to predict the segmentation mask $M^{(c)}$ for class $c$. The entire process can be represented as:
\begin{subequations}\small
    \begin{align}
    &F_I = E_I(I), 
    \label{eq:image_encoder}\\
    &P_s^{(c)}, P_d^{(c)} = E_P(F_I, c) ,
    \label{eq:prompt_encoder}\\
    &M^{(c)} = D_M(F_I, P_p, P_s^{(c)}, P_d^{(c)}) .
    \label{eq:mask_decoder}
    \end{align}
\label{eq:autoMedSAM}
\end{subequations}
\subsection{Diffusion-based Class Prompt Encoder}
\begin{figure}[t]
    \centering
    \includegraphics[width=1\linewidth]{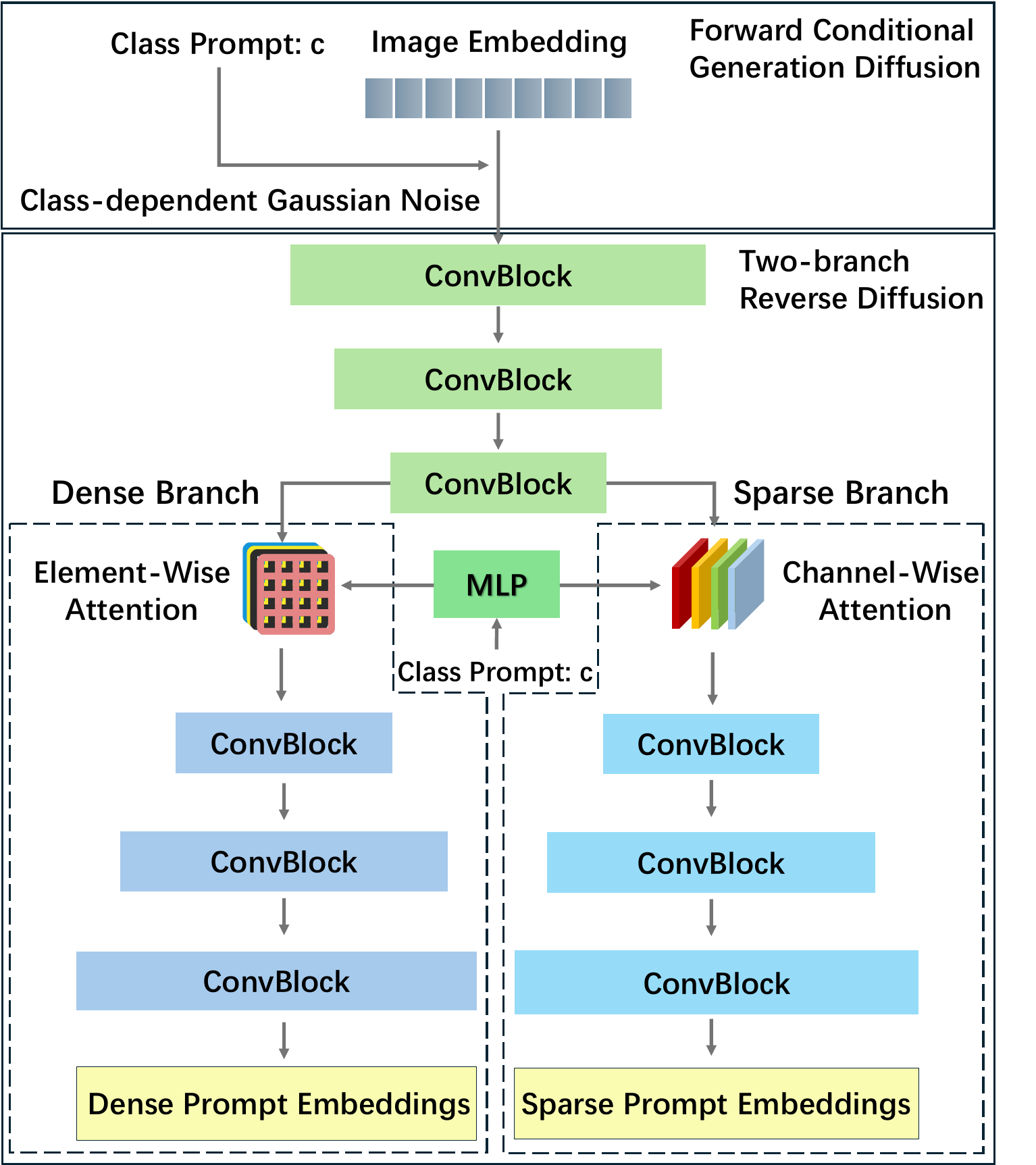}
    \caption{\small Structure of the diffusion-based class prompt encoder. It is designed with an encoder and two independent decoder branches to extract local and global features, based on the practical significance of sparse and dense prompts. The use of prompt classes enables the model to more effectively focus on parts of the input related to specific classes, enhancing its ability to perceive and distinguish class-specific features, thereby improving the controllability and quality of the generation process.}
    \label{fig:overview-diffusion}
    \vspace{-5mm}
\end{figure}

Through an exploration of the SAM mechanism, we found that the goal of the prompt encoder is to generate sparse and dense prompt embeddings. To eliminate the need for manual prompts, we propose a diffusion-based class prompt encoder that integrates the diffusion process with an encoder-decoder framework. Fig.~\ref{fig:overview-diffusion} shows the detailed network structure. As seen, it consists of an encoder and two decoders. The two decoder branches are tasked with generating sparse and dense prompt embeddings, respectively. Additionally, the class prompt not only guides the entire generation process but also ensures that the model's prediction masks carry semantic information.
\subsubsection{Forward Conditional Generation Diffusion}
In the forward diffusion process, the prompt class $c$ is projected and integrated into the noise generation process, enabling the image embeddings to incorporate class information at each step of the diffusion. This approach helps enhance the model’s ability to capture class-specific features when processing inputs with distinct class attributes. Specifically, the class prompt is projected through a linear layer to match the dimensions of the image embedding. The projection process can be represented as:
\begin{equation}
c_{\text{proj}} = W_c c + b_c ,
\label{eq:proj}
\end{equation}
where $ W_c \in \mathbb{R}^{H \times W} $ and $ b_c \in \mathbb{R}^{H \times W} $ are the weight matrix and bias vector of the linear layer. The projected class prompt $ c_{\text{proj}} $ is then reshaped to:
\begin{equation}
c_{\text{expand}} = c_{\text{proj}}.view(B, 1, H, W) \in \mathbb{R}^{B \times 1 \times H \times W} .
\label{eq:reshpe}
\end{equation}
At each time step $ t $, the generated Gaussian noise $ \epsilon_t $ follows a normal distribution with zero mean and variance $ \sigma_t^2 $:
\begin{equation}
\epsilon_t \sim \mathcal{N}(0, \sigma_t^2), \quad \sigma_t = \frac{1}{t+1}
\end{equation}
As the time step $ t $ increases, the noise scale gradually decreases. Finally, the forward diffused embedding $ F_t $ is obtained by adding the image embedding $ F_I $, the projected class prompt $ c_{\text{expand}} $, and the Gaussian noise $ \epsilon_t $ together:
\begin{equation}
F_t = F_I + \epsilon_t + c_{\text{expand}} .
\end{equation}
By conditional generation, we integrate class information into the noise generation process, making the forward diffusion process conditional rather than unconditional. This enhances the controllability of the generation process. This approach enhances feature representation quality and strengthens the model's ability to differentiate class-specific features, thereby improving prompt embedding generation performance.

\subsubsection{Two-branch Reverse Diffusion}
In SAM, dense prompt embeddings capture fine-grained local information specific to a target, while sparse prompt embeddings emphasize capturing broader global features. To distinguish the functional roles of the two embedding types, our diffusion-based class prompt encoder incorporates a single encoder with two independent decoder branches~\cite{ronneberger2015u}, designed to analyze local and global features and produce distinct prompt embeddings. Based on the specific use of dense and sparse prompt embeddings, element-wise attention is applied to the dense prompt branch, while channel-wise attention is used for the sparse prompt branch. Additionally, the model employs skip connections between the encoder and decoder to retain low-level features~\cite{wu2019wider}. During this process, the prompt class $c$ is encoded and combined with embeddings from the encoder, enabling the model to focus on input features relevant to the specific category more effectively. This enhances the model’s ability to perceive and distinguish category-specific features, thereby improving the quality and specificity of the generation process.

Specifically, the diffusion embedding $ F_t \in \mathbb{R}^{B \times C \times H \times W} $ obtained from forward diffusion is fed into the encoder, which captures its features progressively. The encoding process can be represented as:
\begin{equation} 
F_{\text{enc}}^{(l)} = \sigma(W^{(l)} * F_{\text{enc}}^{(l-1)} + b^{(l)}), \quad \forall l = 1, 2, \dots, L
\end{equation}
where $ F_{\text{enc}}^{(l)} $ is the output feature at layer $ l $, $ F_{\text{enc}}^{(0)} = F_t $ represents the diffusion embedding, $ W^{(l)} $ is the weight matrix of the convolution kernel, $ b^{(l)} $ is the bias vector, $ * $ denotes the convolution operation, and $ \sigma $ represents the ReLU activation function. 

After obtaining the encoder's output feature $F_{\text{enc}}^{(l)}$, we code the class prompt $c$ via~\eqref{eq:proj} and~\eqref{eq:reshpe} to align with the feature map's dimensions.
Subsequently, we concatenate $F_{\text{enc}}^{(l)}$ with the post-coding prompt $c_{\text{p}}$ along the channel dimension.
The concatenated feature $ F_{\text{att}}^{(l)} $ contains the image information of the prompt class. It is then passed through the dense prompt embedding branch and the sparse prompt embedding branch separately:

\textbf{Dense Prompt Embedding Branch}. To refine the exploration of local features, we compute the attention weights $A_{\text{dense}}^{(l)}$ using an element-wise convolution operation:
\begin{equation}
A_{\text{dense}}^{(l)} = \sigma(W_{\text{att}}^{(l)} * F_{\text{att}}^{(l)} + b_{\text{att}}^{(l)}) .
\end{equation}
Next, we apply these attention weights to $ F_{\text{enc}}^{(l)} $, automatically focusing on the fine-grained features related to the prompt.
\begin{equation}
F_{\text{dense}}^{(l)'} = F_{\text{enc}}^{(l)} \odot A_{\text{dense}}^{(l)} ,
\end{equation}
where $\odot$ denotes element-wise multiplication. The resulting attention-enhanced feature $F_{\text{enc}}^{(l)'}$ is subsequently concatenated with the corresponding feature from the skip connection and provided as input to the decoder for progressive layer-by-layer decoding:
\begin{equation}
\label{decoding}
P_d^{(c)} = F_{\text{dec}}^{(l-1)} = \sigma(W_{\text{dec}}^{(l)} * F_{\text{dec}}^{(l)} + b_{\text{dec}}^{(l)}) .
\end{equation}
The $ P_d^{(c)} $ obtained from the dense branch contains rich features related to the prompt.

\textbf{Sparse Prompt Embedding Branch}. The sparse prompt embedding necessitates the branch to develop a deeper understanding of global information. To achieve this, we perform global adaptive average pooling on the information from each channel for compression, yielding the global feature $F_{\text{global}}^{(l)}$:
\begin{equation}
F_{\text{global}}^{(l)} = \text{AdaptiveAvgPool2D}(F_{\text{att}}^{(l)}) .
\end{equation}
We then determine the channel attention weights $A_{\text{sparse}}^{(l)}$ via $Sigmoid$ :
\begin{equation}
A_{\text{sparse}}^{(l)} = \text{Sigmoid}(W_{\text{att}}^{(l)} * F_{\text{global}}^{(l)} + b_{\text{att}}^{(l)}) .
\end{equation}
The obtained channel attention weights are subsequently applied to the encoder's output feature to enhance globally relevant class information:
\begin{equation}
F_{\text{sparse}}^{(l)'} = F_{\text{enc}}^{(l)} \otimes A_{\text{sparse}}^{(l)} ,
\end{equation}
where $\otimes$ represents a weighted operation applied to each channel.

After $ F_{\text{sparse}}^{(l)'} $ undergoes the same decoding process as the dense prompt branch, it is transformed in size to obtain the sparse prompt embedding $ P_s^{(c)} $. This embedding provides global features related to the class prompt. Combined with $ P_d^{(c)} $, it meets the requirements of the mask decoder, thereby improving the quality of the generated mask.

\subsection{Uncertainty-aware Joint Optimization}
AutoMedSAM is designed to accommodate various modalities of medical images, which often have significant differences. Moreover, medical images generally have characteristics such as low contrast and high similarity among targets. To enhance segmentation accuracy, the model's optimization process needs to integrate multiple loss functions to improve overall performance. However, as the number of loss functions increases, adjusting their weights becomes increasingly challenging, and these weights are crucial for model optimization as they directly influence the optimization direction~\cite{TsaiUncertainty}. We introduced an uncertainty-aware joint optimization algorithm to adaptively adjust the loss weights. Without the need for any hyperparameters, this strategy autonomously integrates pixel-level, region-level, distribution-level, and other loss functions to comprehensively enhance the model during training.  Specifically, we employed the following loss functions:

\noindent
\textbf{(1) Mean Squared Error (MSE) Loss}: A pixel-based loss that focuses on the difference between predicted and true values for each pixel.

\noindent
\textbf{(2) Cross-Entropy (CE) Loss}: A pixel-based loss that measures the degree of alignment between the predicted class probabilities and the target labels at each pixel.

\noindent
\textbf{(3) Dice Coefficient (DC) Loss}: A region-based metric that measures the overlap between the predicted region and the true region.

\noindent
\textbf{(4) Shape Distance (SD) Loss}: A distribution-based loss that emphasizes the geometric and structural information of the target.

We first generate the sparse prompt embedding $P_{s,i}^{(c)}$ and dense prompt embedding $P_{d,i}^{(c)}$ using the original MedSAM prompt encoder and supervise the generation process of our diffusion-based class prompt encoder using MSE. This can be represented as:
\begin{subequations}\small
    \begin{align}
    &L_{\text{MSE}^{S}} = \frac{1}{n} \sum_{c \in C} \sum_{i=1}^{n} (P_{s,i}^{(c)} - \hat{P_{s,i}^{(c)}})^2 ,
    \label{eq:sparse}\\
    &L_{\text{MSE}^{D}} = \frac{1}{n} \sum_{c \in C} \sum_{i=1}^{n} (P_{d,i}^{(c)} - \hat{P_{d,i}^{(c)}})^2 ,
    \label{eq:dense}
    \end{align}
\label{eq:embedding_loss}
\end{subequations}
where $n$ represents the total number of samples, and $C$ represents the set containing the prompted classes. In this way, our class prompt encoder can quickly leverage the extensive pre-trained knowledge from MedSAM's original prompt encoder. For a given predicted mask and ground truth mask, we use the Dice Coefficient (DC) loss to evaluate their overlap, defined as:
{\small
\begin{equation}
L_{DC} = 1 - \frac{2 \sum_{c \in C} \sum_{i=1}^{n} M_i^{(c)} \hat{M_i^{(c)}}}{\sum_{c \in C} \left( \sum_{i=1}^{n} (M_i^{(c)})^2 + \sum_{i=1}^{n} (\hat{M_i^{(c)}})^2 \right)} .
\end{equation}
}
Next, the classification probability for each pixel is evaluated using Cross-Entropy (CE) loss:
{\small
\begin{align}
L_{CE} = & -\frac{1}{n} \sum_{c \in C} \sum_{i=1}^{n} \Big[ M_i^{(c)} \log(\hat{M_i^{(c)}}) \notag \\
 & + (1 - M_i^{(c)}) \log(1 - \hat{M_i^{(c)}}) \Big] .
\end{align}
}
For each channel $ch$ of each sample, Shape Distance (SD) loss strengthens boundaries and shape consistency by calculating the average difference between the predicted result and the distance transform of the ground truth $D$. The calculation process of $L_{SD}$ is as follows~\cite{Huangshape}:
\begin{subequations}
\begin{align}
    &f_{i, ch} =  \frac{\sum_{h,w} |D({M_{i, ch}^{(c)}}(h,w)) - \hat{M_{i,ch}^{(c)}(h,w)}|}{\sum_{h,w} \hat{M_{i,ch}^{(c)}(h,w)}} , \\
    &L_{SD} = \frac{1}{n \cdot h} \sum_{im=1}^{n} \sum_{ch=1}^{h} f_{i, ch} .
\end{align}
\end{subequations}
Finally, we combine all the losses through an uncertainty-aware framework. It is defined as:
\begin{equation}
L = \sum_{j=1}^{J} \left( \frac{1}{2\lambda_j^2} L_j + \log(1 + \lambda_j^2) \right) ,
\label{eq:aware_loss}
\end{equation}
where $J$ represents the number of loss members involved in the optimization during training, $L_j$ represents all the losses we mentioned above, and $\lambda_j$ are learnable parameters that adjust the contribution of each loss component based on uncertainty.

\section{Experiment}
\subsection{Experimental Settings}
\subsubsection{Datasets}
To evaluate the generalizability of AutoMedSAM, we conducted tests on several commonly used benchmark datasets, including AbdomenCT1K\cite{ma2021abdomenct}, BraTS\cite{menze2014multimodal}, Kvasir-SEG\cite{jha2020kvasir}, and Chest Xray Masks and Labels(Chest-XML)\cite{candemir2013lung}. All data were obtained from the CVPR 2024 Medical Image Segmentation on Laptop Challenge\footnote{https://www.codabench.org/competitions/1847}. These datasets encompass four distinct imaging modalities: CT, MR, endoscopy, and X-ray, covering diverse segmentation targets ranging from organs to lesions. This diversity effectively demonstrates the broad applicability of our method. 
Table~\ref{tab:dataset} provides an overview of the datasets used.
\begin{table}[t]
\caption{Description of the dataset used in this paper.}
\centering
\scalebox{0.8}
{
\begin{tabular}{lll}
\hline
Dataset Name   & Modality   & Segmentation Targets\\ \hline
AbdomenCT-1K\cite{ma2021abdomenct}  & CT(3D)  & Liver, kidneys, pancreas, spleen \\
BraTS\cite{menze2014multimodal}  & MR-FLAIR(3D)    & Brain tumor   \\
Kvasir\cite{jha2020kvasir}   & Endoscopy(2D)   & Polyp \\
Chest Xray Masks and Labels\cite{candemir2013lung} & Chest X-Ray(2D) & Lung \\ 
AMOS\cite{amos} & CT(3D) & Liver, kidneys, pancreas, spleen \\ \hline
\end{tabular}
}
\label{tab:dataset}

\end{table}

\subsubsection{Evaluation Metrics}\label{eval_metrics}
To quantitatively evaluate the segmentation results, we adopted the Dice Similarity Coefficient (DSC) and Normalized Surface Distance (NSD)~\cite{ma2024medsam}.
The DSC is a region-based metric used to evaluate the degree of overlap between the predicted segmentation mask and the expert annotation mask. It is defined as:
\begin{equation}
    DSC(G, S) = \frac{2 |G \cap S|}{|G| + |S|} ,
\end{equation}
where $ G $ and $ S $ denote the ground truth mask and the predicted segmentation mask, respectively. 

The NSD is a boundary-based metric that measures the agreement between the boundaries of the predicted segmentation and the expert annotation, considering a specified tolerance $ \tau $. It is defined as:
\begin{equation}
    NSD(G, S) = \frac{|\partial G \cap B_{\partial S}^{(\tau)}| + |\partial S \cap B_{\partial G}^{(\tau)}|}{|\partial G| + |\partial S|} ,
\end{equation}
where $ B_{\partial G}^{(\tau)} $ and $ B_{\partial S}^{(\tau)} $ denote the border regions around the boundaries of the ground truth and the predicted mask, respectively, within the tolerance $ \tau $. In evaluation, we set the tolerance $ \tau $ to 2. 
For both metrics, a value approaching 1 represents superior segmentation performance, highlighting accurate spatial overlap and boundary consistency with the ground truth annotations.

\subsubsection{Efficient Tuning}
AutoMedSAM demonstrates high training efficiency by employing a selective tuning strategy. During the tuning phase, the large image encoder is kept frozen, while only the diffusion-based prompt encoder and mask decoder parameters are updated. This end-to-end tuning process is conducted under the supervision of the objective defined in~\eqref{eq:aware_loss}, ensuring efficient optimization of the relevant components.

\subsubsection{Implementation Details}
 All experiments were conducted using PyTorch and trained on an NVIDIA RTX A40 GPU. We set the batch size to 16 during training and trained for 100 epochs. The training process utilized the AdamW optimizer with a learning rate of $ lr = 5 \times 10^{-4} $. The optimizer employed hyperparameters $ \beta_1 = 0.9 $, $ \beta_2 = 0.999 $, $ \epsilon = 10^{-8} $. Additionally, a learning rate scheduler was used to reduce the learning rate on a plateau with a factor of 0.9, patience of 5 epochs, and no cooldown period.

\subsubsection{Baseline Methods}
 We compared our method with state-of-the-art medical imaging SAM models, grouped into SAM-Core models (SAM~\cite{SAM}, SAM2~\cite{ravi2024sam}, MedSAM~\cite{ma2024medsam}, Med2d~\cite{cheng2023sam2d}, and U-MedSAM~\cite{wang2024u}) and SAM-Based models (SAMed~\cite{zhang2023SAMed}, H-SAM~\cite{cheng2024H-SAM}, AutoSAM~\cite{hu2023AutoSAM}, and SurgicalSAM~\cite{yue2024surgicalsam}). The SAM-Core Model retains the foundational framework of SAM, requiring manual prompting for operation. The SAM-Based Model builds upon SAM, introducing enhancements that eliminate the need for manual prompting. Notably, SurgicalSAM, like our approach, uses class prompts and prototype contrastive learning to distinguish surgical instruments. To ensure the reliability of the experimental results, all methods were conducted under identical experimental conditions. Besides, we use expert models (i.e., nnU-Net~\cite{nnU-Net}, Swin-Unet~\cite{Swin-unet}, and MedFormer~\cite{Medformer}) as performance benchmarks.


\subsection{Comparing with the Existing Methods}
The interactive segmentation process requires the model to have a multi-objective segmentation capability. Accordingly, we first evaluated the performance of various models on the multi-organ segmentation task using the AbdomenCT1K dataset. The results are shown in Table~\ref{tab:ct}. It can be observed that the performance of different models varies significantly under the DSC and NSD metrics, with AutoMedSAM demonstrating the best overall performance among all methods. Specifically, AutoMedSAM achieves the highest scores on both overall metrics, DSC (94.580\%) and NSD (95.148\%). This improvement is not only reflected in the average performance, but also stands out at the single-organ level(e.g., Pancreas and Kidney). The performance gain can be attributed to the proposed diffusion-based class prompt encoder and uncertainty-aware joint optimization strategy, which refines the structure and detail of the prompt class through the diffusion process and injects semantic information into the prediction pipeline. 
In contrast, although SAM-Core models (e.g., MedSAM with an NSD of 92.969\%) offer stable performance due to manual prompting, they suffer from inefficiency in practical use. This limitation makes them less suitable for time-sensitive clinical scenarios where fast segmentation of complex medical images is essential. SAM-Based models, while capable of automatic prompting, generally perform worse in terms of segmentation accuracy. All models in this category, except AutoMedSAM, fall below the SAM-Core baseline. Traditional specialist models deliver relatively balanced performance and even excel in specific organs (e.g., nnU-Net with a DSC of 96.317\% on the right kidney), yet they lack the generalization and flexibility needed for interactive segmentation tasks, limiting their broader applicability.

\begin{table*}[t]
\caption{Comparative Results on the AbdomenCT-1K Dataset. The SAM-Core Model retains the framework of SAM, requiring manual prompting for operation. The SAM-Based Model builds upon SAM, introducing enhancements that eliminate the need for manual prompting. L. and R. stand for left and right. $\uparrow$ means higher is better. The best results are shown in \textbf{Bold}.}
\centering
\scalebox{0.78}{
\begin{tabular}{|l|l|cc|ccccc|ccccc|}
\hline
\multirow{3}{*}{Method   Category} & \multirow{3}{*}{Method} & \multirow{3}{*}{DSC(\%)$\uparrow$} & \multirow{3}{*}{NSD(\%)$\uparrow$}  & \multicolumn{10}{c|}{Organ}  \\ \cline{5-14} 
 && &  & \multicolumn{5}{c|}{DSC(\%)$\uparrow$}  & \multicolumn{5}{c|}{NSD(\%)$\uparrow$}   \\ \cline{5-14} 
 && &  & Liver    & Spleen  & Pancreas & L.Kidney & R. Kidney& Liver   & Spleen   & Pancreas & L. Kidney& R. Kidney\\ \hline

\multirow{3}{*}{Specialist   Model} & \multicolumn{1}{l|}{nnU-Net~\cite{nnU-Net}}   & 93.196     & \multicolumn{1}{c|}{92.621} & 96.187     & 96.130     & 82.095     & 95.249     & \multicolumn{1}{c|}{\textbf{96.317}} & 90.962     & 97.573     & 83.351     & 94.103     & \multicolumn{1}{c|}{\textbf{97.118}} \\
\multicolumn{1}{|l|}{}        & \multicolumn{1}{l|}{Swin-Unet~\cite{Swin-unet}} & 87.996     & \multicolumn{1}{c|}{86.215} & 97.057     & 94.790     & 68.878     & 88.874     & \multicolumn{1}{c|}{90.382}          & \textbf{96.132}      & 97.126     & 65.835     & 80.419     & \multicolumn{1}{c|}{91.563}          \\
\multicolumn{1}{|l|}{}        & \multicolumn{1}{l|}{MedFormer~\cite{Medformer}} & 92.479     & \multicolumn{1}{c|}{88.452} & 96.883     & 94.816     & 80.942     & 94.861     & \multicolumn{1}{c|}{94.894}          & 95.491     & 90.733     & 85.066     & 87.668     & \multicolumn{1}{c|}{83.300}          \\  \hline
 
\multirow{5}{*}{SAM-Core Model}       & SAM\cite{SAM} & 89.790   & 83.940    & 92.117   & 93.766  & 72.694   & 95.254   & 95.121   & 75.376  & 90.253   & 69.687   & 92.385   & 91.998   \\
 & SAM2\cite{ravi2024sam}       & 90.191  & 85.137   & 93.815   & 95.391  & 70.793   & 95.623   & 95.334   & 81.804  & 94.149   & 62.519   & 93.860    & 93.354   \\
 & MedSAM\cite{ma2024medsam}     & 93.505      & 92.969       & 96.836       & \textbf{97.120} & 81.648       & 96.121       & 95.800 & 91.908      & 98.807       & 83.845       & 95.374       & 94.913       \\
 & Med2d\cite{cheng2023sam2d}      & 83.840   & 79.347   & 93.562   & 93.371  & 58.168   & 87.081   & 87.020    & 80.020   & 89.751   & 67.776   & 79.624   & 79.566   \\
 & U-MedSAM\cite{wang2024u}   & 92.979  & 91.158   & 96.606   & 96.799  & 79.841   & 95.906   & 95.742   & 90.469  & 98.327   & 76.932   & 95.258   & 94.804   \\ \hline
\multirow{5}{*}{SAM-Based model}   & SAMed\cite{zhang2023SAMed}      & 81.329  & 78.504   & 97.132   & 96.838  & 76.512   & 67.904   & 68.259   & 92.435  & 95.390    & 77.635   & 63.342   & 63.716   \\
 & H-SAM\cite{cheng2024H-SAM}      & 83.018  & 78.852   & 96.678   & 96.447  & 77.337   & 71.856   & 72.774   & 91.641  & 95.392   & 78.946   & 63.894   & 64.388   \\
 & AutoSAM\cite{hu2023AutoSAM}    & 82.258  & 76.305   & 96.326   & 95.625  & 72.555   & 73.206   & 73.578   & 89.899  & 91.574   & 72.391   & 63.589   & 64.073   \\
 & SurgicalSAM\cite{yue2024surgicalsam}& 75.505  & 70.119   & 96.054   & 94.255  & 75.621   & 54.915   & 56.683   & 87.399  & 92.303   & 73.985   & 48.386   & 48.524   \\ 
 & \textbf{AutoMedSAM(Ours)}   & \textbf{94.580} & \textbf{95.148} & \textbf{97.467} & 96.958      & \textbf{86.061} & \textbf{96.291} & 96.121 & 95.030 & \textbf{98.911} & \textbf{88.903} & \textbf{96.585} & 96.309 \\ \hline
\end{tabular}
}
\label{tab:ct}
\end{table*}

\begin{figure}[t]
    \centering
    \includegraphics[width=1\linewidth]{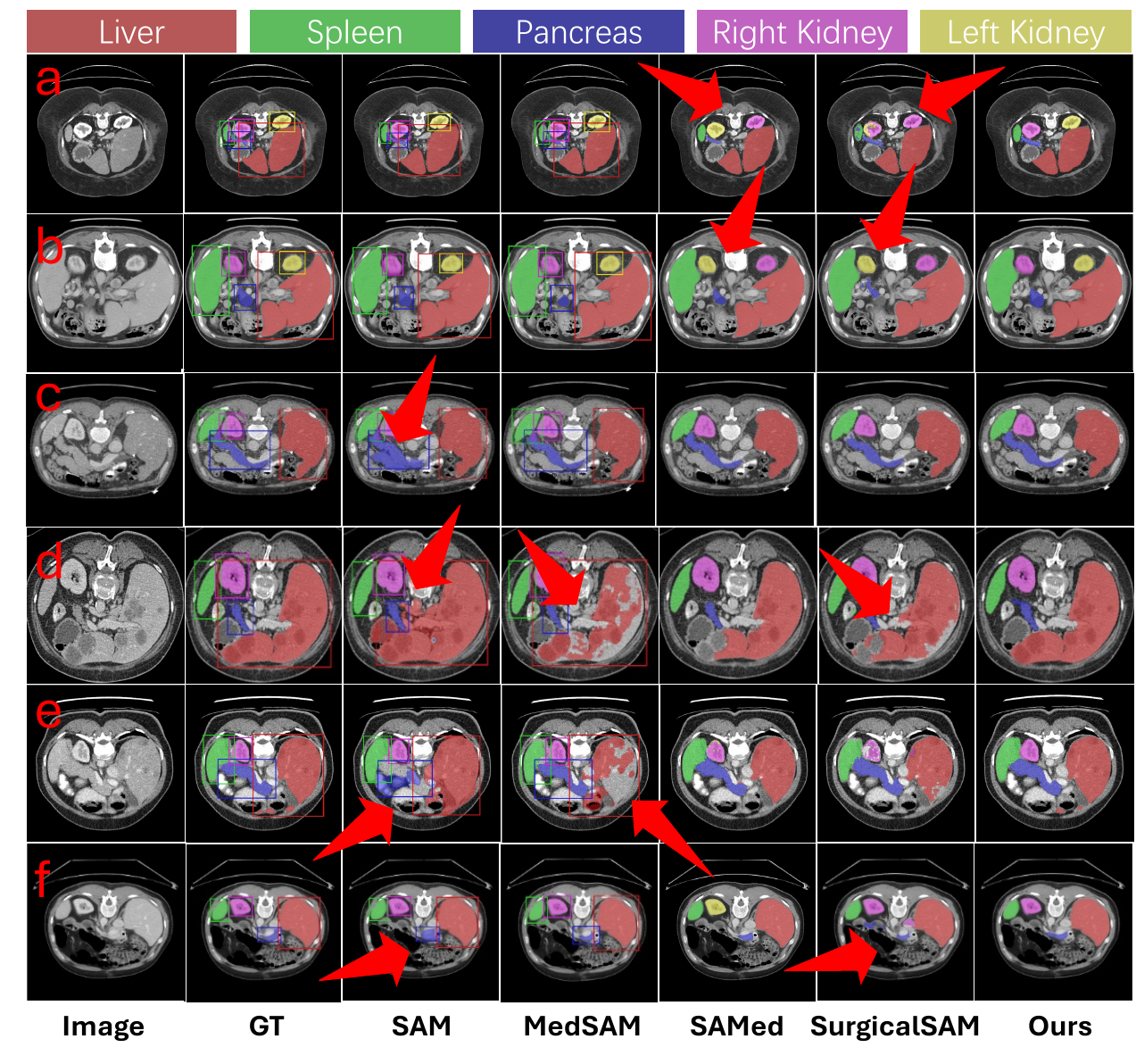}
    \caption{\small The qualitative results of AutoMedSAM and other comparison models on AbdomenCT-1K. The bounding box represents the input prompt.}
    \label{fig:ct_vis}
\end{figure}

As shown in Table~\ref{tab:ct}, the segmentation performance of the tested models varies minimally for the liver, spleen, and pancreas but significantly deteriorates for kidney segmentation. To investigate this, we visualized the segmentation results (shown in Fig.~\ref{fig:ct_vis}). Fig.~\ref{fig:ct_vis} (a) and (b) illustrate that while SAM-Based models generally perform well in boundary recognition, they confuse the left and right kidneys. This is due to the strong symmetry and high morphological similarity of the left and right kidneys, and the SAM-Core model addresses this challenge by providing location information of the target organs through manual prompting. However, the automatic prompting mechanism of the SAM-Based model lacks this information. This is why their performance drops sharply. However, during manual prompting, bounding boxes may unavoidably include other organs. Since SAM-Core models lack semantic information during prediction, they cannot accurately identify specific organs, leading to segmentation errors. As shown in Fig.~\ref{fig:ct_vis} (c) and (d), both SAM and MedSAM mistakenly identified other tissues to varying degrees. Additionally, even with minimal redundancy within bounding boxes, SAM models tend to segment features that appear prominent, which degrades mask quality (shown in Fig.~\ref{fig:ct_vis} (e) and (f)). In contrast, our method incorporates class-based prompts to introduce semantic information into predictions, effectively mitigating this issue.

\begin{figure}[ht]
    \centering
    \includegraphics[width=1\linewidth]{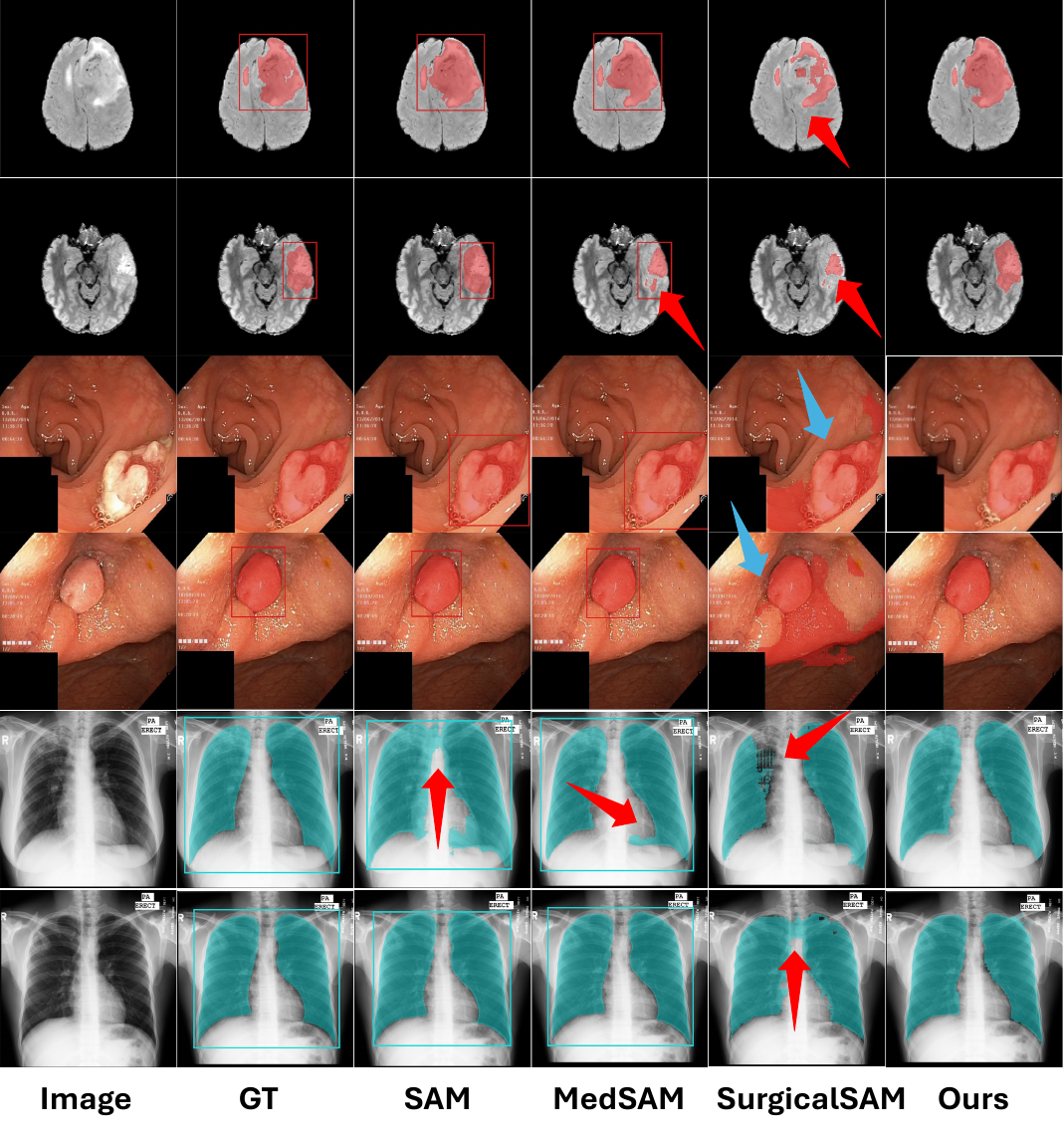}
    \caption{\small The qualitative analysis results of AutoMedSAM and other comparison models on BraTS, Kvasir-SEG, and Chest-XML.}
    \label{fig:xray_mr_endos_vis}
\end{figure}

To evaluate segmentation accuracy independently, we conducted additional experiments on the BraTS, Kvasir-SEG, and Chest-XML datasets. In BraTS and Kvasir-SEG, the targets (tumor and polyp) are singular but structurally complex, with diverse shapes and ambiguous boundaries. In Chest-XML, overlapping structures like ribs around the lungs and the large target area introduce significant challenges. These datasets provide diverse difficulties suitable for evaluating segmentation performance. The results and visualizations are shown in Table~\ref{tab:single-object} and Fig.~\ref{fig:xray_mr_endos_vis}, respectively. 

\begin{table*}[ht]
\caption{Comparative Results on the BraTS, Kvasir-SEG and Chest-XML Datasets}
\centering
\scalebox{1}{
\begin{tabular}{|l|l|cccccc|}
\hline
\multirow{3}{*}{Method   Category} & \multirow{3}{*}{Method} & \multicolumn{6}{c|}{Dataset(Lesion/Organ)}     \\ \cline{3-8} 
 &        & \multicolumn{2}{c|}{BraTS(Tumor)}& \multicolumn{2}{c|}{Kvasir-SEG(Polyp)}  & \multicolumn{2}{c|}{Chest-XML(Lung)} \\ \cline{3-8} 
 &        & DSC(\%)$\uparrow$  & \multicolumn{1}{c|}{NSD(\%)$\uparrow$}  & DSC(\%)$\uparrow$  & \multicolumn{1}{c|}{NSD(\%)$\uparrow$}  & DSC(\%)$\uparrow$      & NSD(\%)$\uparrow$      \\ \hline
\multirow{3}{*}{Specialist   Model} & \multicolumn{1}{l|}{nnU-Net~\cite{nnU-Net}}   & 87.860     & \multicolumn{1}{c|}{87.611} & 90.854     & \multicolumn{1}{c|}{92.692} & 95.972     & \multicolumn{1}{c|}{96.457} \\
\multicolumn{1}{|l|}{}  & \multicolumn{1}{l|}{Swin-Unet~\cite{Swin-unet}} & 86.236     & \multicolumn{1}{c|}{84.235} & 88.465     & \multicolumn{1}{c|}{91.496} & 94.285     & \multicolumn{1}{c|}{94.840} \\
\multicolumn{1}{|l|}{}  & \multicolumn{1}{l|}{MedFormer~\cite{Medformer}} & 89.251     & \multicolumn{1}{c|}{89.398} & 91.647     & \multicolumn{1}{c|}{94.649} & 96.204     & \multicolumn{1}{c|}{96.694} \\  \hline       
\multirow{5}{*}{SAM-Core Model}         & SAM\cite{SAM}    & 69.667& \multicolumn{1}{c|}{42.112}& 92.749& \multicolumn{1}{c|}{95.384}& 94.326   & 94.878   \\
 & SAM2\cite{ravi2024sam}   & 77.061& \multicolumn{1}{c|}{57.808}& 94.013& \multicolumn{1}{c|}{96.282}& 95.423   & 95.975   \\
 & MedSAM\cite{ma2024medsam} & 89.568& \multicolumn{1}{c|}{89.517}       & 95.803       & \multicolumn{1}{c|}{97.829}       & 95.864     & 96.354    \\
 & Med2d\cite{cheng2023sam2d}  & 63.327& \multicolumn{1}{c|}{73.528}& 81.609& \multicolumn{1}{c|}{87.202}& 92.805   & 93.412   \\
 & U-MedSAM\cite{wang2024u} & 89.616       & \multicolumn{1}{c|}{88.793}& 95.007& \multicolumn{1}{c|}{97.091}& 96.813   & 97.310    \\ \hline
\multirow{5}{*}{SAM-Based model}   & SAMed\cite{zhang2023SAMed}  & 89.902& \multicolumn{1}{c|}{89.097}& 86.393& \multicolumn{1}{c|}{88.255}& 94.066   & 94.600     \\
 & H-SAM\cite{cheng2024H-SAM}  & 90.615& \multicolumn{1}{c|}{90.288}& 88.010& \multicolumn{1}{c|}{90.084}& 92.264   & 92.978   \\
 & AutoSAM\cite{hu2023AutoSAM} & 90.240& \multicolumn{1}{c|}{89.186}& 87.045& \multicolumn{1}{c|}{89.288}& 89.616   & 90.516    \\
 & SurgicalSAM\cite{yue2024surgicalsam}   & 80.373& \multicolumn{1}{c|}{75.654}& 78.831& \multicolumn{1}{c|}{80.684}& 91.406   & 92.063   \\
 & \textbf{AutoMedSAM(Ours)}& \textbf{91.057} & \multicolumn{1}{c|}{\textbf{92.661}} &  \textbf{96.828} & \multicolumn{1}{c|}{\textbf{98.729}} & \textbf{96.941}         &  \textbf{97.367} \\ \hline
\end{tabular}
}
\label{tab:single-object}
\end{table*}

As shown in Table~\ref{tab:single-object}, AutoMedSAM achieved superior performance across all tasks, with DSC and NSD scores of 96.828\% and 98.729\% for polyp segmentation, demonstrating its adaptability to complex medical environments. Conversely, when SAM-Core models lose their advantage of manual prompts, their performance declines significantly. For example, in BraTS, SAM achieved only 69.667\% DSC and 42.112\% NSD for tumor segmentation, indicating reasonable overlap with ground truth but poor boundary recognition. Similar issues were observed in lung segmentation, as shown in the last two subfigures of Fig.~\ref{fig:xray_mr_endos_vis}, where SAM models produced masks containing excessive non-target tissues. 
In conclusion, the experimental results confirm that the proposed AutoMedSAM effectively adapts to various medical modalities, delivering accurate segmentation masks and achieving the best metrics across all datasets.

\subsection{Cross-Dataset Generalization}
To verify the cross-dataset and cross-modality generalization of AutoMedSAM, we trained the model on AbdomenCT-1K and BraTS-FLAIR, then tested it on AMOS and BraTS-T1CE. The results are shown in Table~\ref{tab:cross_data}, where only the organ or lesion classes shared by both datasets are considered. Our method holds significant advantages compared to SurgicalSAM, which also employs class indexes as prompts. Notably, when generalizing from BraTS-FlAIR to BraTS-T1CE, we achieve a large improvement of 15.67\% in the DSC. This demonstrates that, compared to SurgicalSAM's prototype-based contrastive learning, AutoMedSAM achieves stronger generalization ability by capturing more fine-grained features of organs or lesions through the diffusion process.

\begin{table}[h]
\caption{Cross-Dataset Generalization.}
\scalebox{0.9}{
\begin{tabular}{lllcc}
\hline
Training    & Testing   & Method    & DSC(\%)$\uparrow$   & NSD(\%)$\uparrow$   \\ \hline
\multirow{2}{*}{AbdomenCT-1K} & \multirow{2}{*}{AMOS}       & SurgicalSAM         & 56.930 & 64.451          \\
  & & \textbf{AutoMedSAM} & \textbf{71.141} & \textbf{77.282} \\ \hline
\multirow{2}{*}{BraTS-FLAIR}  & \multirow{2}{*}{BraTS-T1CE} & SurgicalSAM         & 36.843          & 39.825          \\
  & & \textbf{AutoMedSAM} & \textbf{52.513} & \textbf{48.101} \\ \hline
\end{tabular}
}
\label{tab:cross_data}
\end{table}

\subsection{Ablation Study} 
\subsubsection{Effects of Prompt Branch}
The diffusion-based prompt encoder structure shows that we designed two branches to enhance its performance. To determine their impact on the final segmentation accuracy, we conducted experiments on each branch separately (shown in Table~\ref{tab:branch}). During the experiments, one branch was deactivated when the other was activated. The missing prompt embeddings are initialized in accordance with the approach adopted by MedSAM. Table~\ref{tab:branch} shows that AutoMedSAM achieves optimal performance when both branches are activated simultaneously. When the Sparse branch is activated, the NSD decreases by 4.616\%, whereas activating the Dense branch results in a 1.09\% decrease. This demonstrates that the Dense branch plays a complementary role in enhancing boundary details. And the embeddings from both branches mutually reinforce each other, ultimately contributing to optimal prompt.

\begin{table}[ht]
    \centering
    \caption{Ablation study on the prompt branch using the AbdomenCT-1K dataset.}
    \label{tab:branch}
    \begin{tabular}{ccc}
    \hline
        Prompt Branch & DSC(\%)$\uparrow$ & NSD(\%)$\uparrow$\\ \hline
        Dense Branch & 93.752 & 94.058\\
        Sparse Branch & 93.256 & 90.532\\
        Dense + Sparse Branch(Ours) & \textbf{94.580} & \textbf{95.148}\\
    \hline
    \end{tabular}
    
\end{table}

\subsubsection{Effects of Diffusion Processing}
Following the characteristics of diffusion models, AutoMedSAM incorporates class prompts conditionally during the forward diffusion process, enabling the model to learn to emphasize target-specific features across varying noise levels, thereby enhancing its representation of complex anatomical structures. We performed ablation experiments to verify the diffusion process's effect on the prompts. Table~\ref{tab:diffusion} shows the experimental results. From the table, it can be noticed that when the diffusion process is disabled, the DSC and NSD are reduced by 5.379\% and 5.173\% respectively. This proves that diffusion processing is the key to AutoMedSAM's robust generalization in complex clinical settings.


\begin{table}[h]
    \centering
    \caption{Ablation study on the diffusion processing using the Chest-XML dataset.}
    \label{tab:diffusion}
    \begin{tabular}{ccc}
    \hline
        Diffusion Processing & DSC(\%)$\uparrow$ & NSD(\%)$\uparrow$\\ \hline
        \xmark & 91.562 & 92.194\\
        \checkmark & \textbf{96.941} & \textbf{97.367} \\
    \hline
    \end{tabular}
    
\end{table}

\subsubsection{Effects of Uncertainty-aware Joint Optimization}
\begin{table}[t]
\caption{Ablation study on the Uncertainty-aware joint optimization using the Chest-XML dataset.}
\centering
\scalebox{0.95}{
\begin{tabular}{ccccccc}
\hline
Joint Optimization & CE & Dice & SD & MSE & DSC(\%)$\uparrow$& NSD(\%)$\uparrow$\\ \hline
\checkmark    &  \checkmark  &  \checkmark    &  \checkmark  & \xmark   & 92.573  & 93.281  \\
\checkmark    &  \checkmark  &  \checkmark    & \xmark  &  \checkmark   & 95.263  & 95.800\\
\checkmark    &  \checkmark  & \xmark    &  \checkmark  &  \checkmark   & 94.452  & 95.208  \\
\checkmark    &  \xmark  &  \checkmark    &  \checkmark  &  \checkmark   & 93.570   & 94.060   \\
\xmark    &  \checkmark  &  \checkmark    &  \checkmark  &  \checkmark   & 93.359  & 94.113  \\
 \checkmark    &  \checkmark  &  \checkmark    &  \checkmark  &  \checkmark   & \textbf{96.941} & \textbf{97.367} \\ \hline
\end{tabular}
}
\label{tab:joint}
\end{table}
To evaluate the contribution of the proposed uncertainty-aware joint optimization with each loss, we conducted a series of ablation experiments. As shown in Table~\ref{tab:joint}, we progressively disabled the different loss components (e.g., CE, DC, SD, and MSE), and observed the resulting changes in segmentation performance. The full strategy with all four losses included in the uncertainty-aware weighting framework achieved the highest performance with DSC and NSD metrics of 96.941\% and 97.367\%, respectively. When the uncertainty weighting was removed and fixed weights were used, the DSC and NSD metrics decreased by 3.582\% and 3.254\%, respectively, indicating that the adaptive weighting mechanism is very effective in reconciling conflicting loss contributions. Among the individual loss components, removing the MSE loss (used to transfer MedSAM pre-training knowledge) resulted in the most significant decrease (from 96.941\% to 92.573\%), indicating its important role in knowledge transfer. Disabling CE, DC, or SD loss also led to significant degradation, highlighting the complementary nature of pixel-based, region-based, and distribution-based targets when capturing fine anatomical structures. Overall, uncertainty-aware joint optimization can effectively balance multiple loss functions, and improve segmentation accuracy and generalization.

\subsection{Limitations of Manual Prompts} \label{sec:manual}

\begin{figure}[t]
    \centering
    \includegraphics[width=1\linewidth]{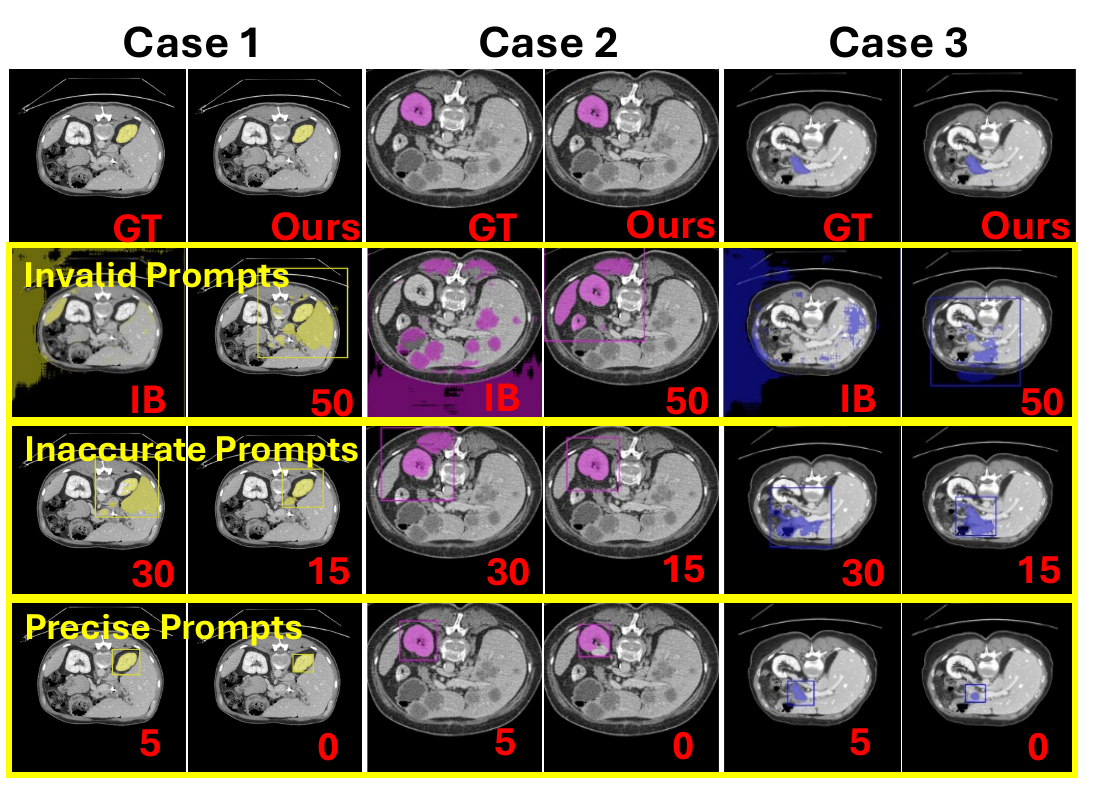}
    \caption{\small The effect of different sized prompt boxes on segmentation masks. The marked numbers indicate the offset pixel size of the prompt box. IB represents the image boundary.}
    \label{fig:prompt_vis}
\end{figure}

As discussed in Sec.~\ref{sec:introduction}, manually sketching an explicit prompt significantly constrains the segmentation accuracy of the model. To further investigate this phenomenon, we explored the impact of prompt boxes with varying offsets on the segmentation results. The experimental results indicate that segmentation accuracy improves as the boundaries of the prompt box approach the target object(shown in Table~\ref{tab:medsam}). However, the accuracy is not the best when the box aligns perfectly with the target boundary. Fig.~\ref{fig:prompt_vis} illustrates the segmentation masks under different prompt box configurations. The figure reveals that overly large prompt boxes include multiple segmentable objects, leading to misidentification by the SAM-core model, as it struggles to determine the specific organ to segment. Conversely, overly small prompt boxes drive the model to search for deeper internal differences within the box, which reduces segmentation accuracy. To address these challenges, we proposed the class prompt method, which incorporates semantic information into the segmentation process. This approach eliminates errors caused by unstable manual prompts, simplifies the segmentation procedure, and enhances the model's robustness.

\begin{table}[ht]
\caption{The impact of boundary box prompts with different accuracies on MedSAM performance. The experimental results were tested on the AbdomenCT1K dataset.}
\centering
\begin{tabular}{lll}
\hline
Box Offset(pixel) & DSC(\%)$\uparrow$         & NSD(\%)$\uparrow$         \\ \hline
0 & 91.301          & 89.816          \\
5 & \textbf{93.505} & \textbf{92.969} \\
15       & 81.714          & 64.624          \\
30       & 48.310 & 28.924          \\
50       & 24.432          & 16.525          \\
Image Boundary     & 2.359 & 2.366 \\ \hline
\end{tabular}
\label{tab:medsam}
\end{table}

\section{Conclusion}
In this paper, we present AutoMedSAM, an end-to-end interactive model for medical image segmentation, designed to address the limitations of manual prompting and the lack of semantic annotation in SAM. By introducing a diffusion-based class prompt encoder, AutoMedSAM eliminates the need for labor-intensive prompts (e.g., points, boxes, or scribbles), and instead utilizes class indices to generate prompt embeddings. The diffusion process conditionally integrates class prompts throughout the generation of embeddings, progressively enhancing the model’s ability to capture the structural and semantic characteristics of organs or lesions, thereby improving the controllability and quality of the generated prompts. The dual-branch architecture further refines both local and global features, boosting segmentation detail and stability. Additionally, we propose an uncertainty-aware joint optimization strategy that adaptively integrates pixel-based, region-based, and distribution-based losses, improving generalization and robustness across diverse medical modalities. Extensive experiments demonstrate that AutoMedSAM achieves state-of-the-art performance, and also shows strong cross-dataset generalization. With the help of AutoMedSAM, doctors and researchers can quickly and accurately analyze medical images in time-sensitive clinical settings, enabling more timely diagnosis and treatment. In the future, we will optimize it on larger-scale medical datasets to enhance its practicality and clinical applicability.

\section*{References}
\bibliographystyle{IEEEtran.bst}
\bibliography{manuscript}

\end{document}